\documentclass[prl,twocolumn,preprintnumbers,amsmath,amssymb]{revtex4}

\usepackage{graphicx}

\begin{document}
\title{Evolution from a molecular Rydberg gas to an ultracold plasma in a seeded supersonic expansion of NO}

\author{J. P. Morrison}
\author{C. J. Rennick}
\author{J. S. Keller}
\altaffiliation{Permanent address: Department of Chemistry, Kenyon College,Gambier, Ohio 43022-9623}
\author{E. R. Grant}
\altaffiliation{Author to whom correspondence should be addressed. Electronic mail:
edgrant@chem.ubc.ca}
\affiliation{Department of Chemistry, University of British Columbia, Vancouver, BC V6T 1Z3, Canada}

\date{\today}

\begin{abstract}
We report the spontaneous formation of a plasma from a gas of cold Rydberg molecules.  Double-resonant laser excitation promotes nitric oxide, cooled to 1 K in a seeded supersonic molecular beam, to single Rydberg states extending as deep as 80 cm$^{-1}$ below the lowest ionization threshold.  The density of excited molecules in the illuminated volume is as high as 1 x 10$^{13}$ cm$^{-3}$.  This population evolves to produce prompt free electrons and a durable cold plasma of electrons and intact NO$^{+}$ ions.  

\end{abstract}


\maketitle

Plasmas are usually hot, and thermal kinetic energy dominates in a binary-collision, gas-like fluid dynamics.  However, under extreme conditions, many-body charged-particle interactions can exceed thermal energies and give rise to liquid- or solid-like spatial correlations.  The degree of correlation ($\Gamma$) depends on the translational energy of the ions compared with their Coulomb repulsion, which varies with plasma density as, $\Gamma  = q^{2 }/4\pi \varepsilon _0 akT$, where $q$ is the charge, and $a$, the Wigner-Seitz radius, is related to the particle density, $\rho$, by, $4/3~\pi a^3  = 1/\rho $  \cite{Ichimaru}.  Strongly correlated plasmas ($\Gamma >$ 1) are thought to exist at the center of heavy stars and under the conditions necessary for inertial confinement fusion.

Recent techniques for cooling distributions of atoms to ultracold temperatures have made it possible to approach the conditions of strong coupling in highly rarified systems in the laboratory \cite{Killian:2008}.  Ultracold plasmas integrate atomic and mesoscopic domains, and thereby open the many-body multiscale heart of the problem to fundamental experimental and theoretical study \cite{Weidemuller_many, Robicheaux_coherent, Murillo,Ates,Raithel_ponder,Tanner}.  

Normal plasmas contain molecules as well as atoms, and molecular degrees of freedom affect their properties.  Similarly, important questions can be raised concerning the role of molecular characteristics, such as rotational relaxation and cation-electron dissociative recombination, in directing the dynamics by which a gas of Rydberg molecules might evolve to form an ultracold plasma.  However, a molecular ultracold plasma has yet to be produced for study.  

Advances in research on atomic Rydberg gases and ultracold plasmas owe a great deal to the development of technology for producing cold atoms and atomic condensates in a magneto-optical trap (MOT) \cite{Killian, Feldbaum, AWalzFlannigan, Li}.  Rotational cycling prevents the laser deceleration of molecules, but one can routinely form dense molecular samples, cooled to temperatures of 1 K and less in seeded supersonic expansions.    

While the molecules entrained in a supersonic beam are much warmer than the micro-Kelvin initial conditions typical in an atom trap, the plasma created in a MOT undergoes substantial excursions in temperature caused by disorder-induced heating followed by electron evaporation and Coulomb expansion \cite{Roberts, Fletcher}.  

By comparison, at a realistic excited-state density of 10$^{12}$ cm$^{-3}$, electrons with a temperature as high as T$_{e}$ = 100 K in the plasma formed by a Rydberg gas in the moving frame of a 1 K supersonic expansion would exhibit a correlation parameter approaching $\Gamma_{e}$ = 0.3.  The cation correlation in such a plasma could exceed 5 \cite{Killian}.  The evaporation of a carrier gas (typically He) presents a cooling mechanism not available in a MOT, improving the prospects for higher correlation.  

The realization of a cold molecular plasma under such conditions would represent a significant development.  A great many small molecules can be seeded in free jets, and this medium would thus represent an important new platform for the broader study of Rydberg gas to ultracold plasma dynamics.  Rydberg gases of alkali atoms have been known for some time to form plasmas under the warmer conditions of an effusive beam  \cite{Vitrant}.  But, for molecular systems, there remain the important questions of predissociation and dissociative recombination:  Can a molecular ultracold plasma form, and if so, how long can it survive?  

In results reported here, we establish that excited NO molecules in ensembles of photoselected Rydberg states lying more than 80 cm$^{-1}$ below the lowest ionization threshold can interact in a supersonic molecular beam to produce a cold plasma that is long lived and remarkably durable.  

In our experiment, a pulsed jet of NO, which is seeded at 10 percent in He at a backing pressure of 5 atm, expands through a 0.5 mm diameter nozzle.  Two cm downstream, an electroformed Ni skimmer selects the 1 mm diameter core of this free jet to form a differentially pumped supersonic molecular beam.  

Flow-field models for such expansions define a quantity, $S^{\infty}_{\|}$, which represents the ratio of the mean terminal flow velocity, $\bar{V}^{\infty}_{\|}$~(defined simply by the source temperature, the average molecular weight and the heat capacity ratio of the expanding mixture) to the most probable longitudinal velocity in the moving frame of the molecular beam, viz. \cite{Miller}:
\[
S^{\infty}_{\|} = \bar{V}^{\infty}_{\|}/\left(\frac{2kT^{\infty}_{\|}}{m}\right)^{\frac{1}{2}},
\]

\noindent Here we can understand $T^{\infty}_{\|}$ as the second moment of the longitudinal velocity distribution function under the terminal conditions of free molecular flow.  

The present stagnation pressure and nozzle diameter predict a seeded jet in which the He carrier has a terminal speed ratio of about 90, giving rise to a parallel temperature of approximately 700 mK for its NO component.  

Translational cooling of the NO is accompanied by rotational cooling as well, but, because rotational relaxation occurs with a smaller collision cross section, equilibration with translational cooling ceases before the onset of molecular flow, with the result that the terminal rotational temperature, T$^{\infty}_{R}$, can be expected to exceed $T^{\infty}_{\|}$.

The centerline intensity for an ideal isentropic expansion relates proportionally to the nozzle flow rate by:
\[
I_{0} = \frac{\kappa}{\pi}F(\gamma) n_{0}  \sqrt{\frac{2kT_{0}}{m}} \left(\frac{\pi d^{2}}{4}\right),
\]

\noindent where $\kappa$ and $F(\gamma)$ are peaking factor and heat capacity function, which for our conditions have values of 1.98 and 0.513 respectively  \cite{Beijerinck}.  n$_{0}$ and $T_{0}$ are the stagnation density and temperature, and d is the nozzle diameter.  For a pure He beam under present conditions, this approximation predicts I$_{0}$ = 9 x 10$^{21}$ particles s$^{-1}$ s$r^{-1}$.  Allowing for attenuation by the skimmer, this flux yields an instantaneous density  of 4.8 x 10$^{14}$ particles cm$^{-3}$ at a distance 10.5 cm from the nozzle.  

Addressing a small volume element at this distance, our experiment excites entrained NO molecules by the sequential application of unfocused 5 ns pulses, $\omega_{1}$ and $\omega_{2}$, from two frequency-doubled Nd:YAG-pumped dye lasers.  First-photon absorption in the region of 226 nm promotes transitions in the A $^{2}\Sigma^{+}$ - X $^{2}\Pi$~ $\gamma$~(0,0) system.  

At laser pulse energies greater than 500 $\mu$J, a scan of $\omega_{1}$ produces an electron signal corresponding to the resonant two-photon ionization spectrum of this band system.  From the line intensities in such scans, we can estimate the population of ground-state rotational levels, from which we can determine T$^{\infty}_{R}$.  For the conditions described above we typically observe a rotational temperature in the range of 3 K.  

Tuning to the specific transition from X $^{2}\Pi ~v'' = 0, N'' = 1 ~(J'' = \frac{1}{2}$) to A $^{2}\Sigma^{+} ~v' = 0, N' = 0 ~(J' = \frac{1}{2}$), we decrease the power of $\omega_{1}$ until the electron signal from one-color resonant two-photon ionization disappears.  At this point, the A $^{2}\Sigma^{+}$ - X $^{2}\Pi$ first step is still saturated, and we overlap a second laser pulse, $\omega_{2}$, to further excite these photoselected NO molecules.  Scanning $\omega_{2}$ over the range from 30,400 to 30,530 cm$^{-1}$ populates high Rydberg states in series converging to the lowest ionization threshold of NO.  

To estimate the absolute density of excited state molecules produced under our conditions, we scale the total density at the interaction distance by the seeding ratio (0.1), the rotational fraction in $J'' = \frac{1}{2}$ at 3 K (0.87), and two factors of 0.5 for saturated steps of $\omega_{1}$ and $\omega_{2}$ excitation.  From this, we obtain an approximate density of high-Rydberg NO molecules of 1 x 10$^{13}$ cm$^{-3}$.  

\begin{figure}
\includegraphics[width=\columnwidth]{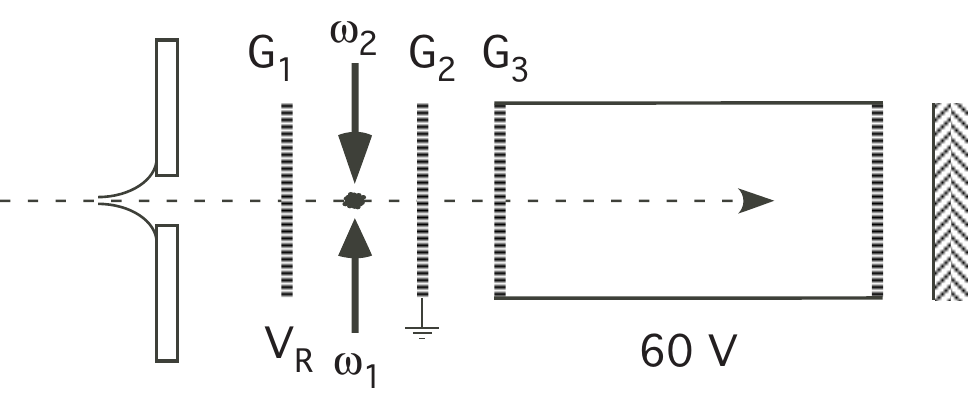}
\caption{Schematic diagram showing the molecular beam flight path from a differentially pumped source chamber through a skimmer to enter a mu-metal shielded system of three grids ending in a flight tube capped by a microchannel plate detector.}
\label{Figure_1.pdf}
\end{figure}

\begin{figure}
\includegraphics[width=\columnwidth]{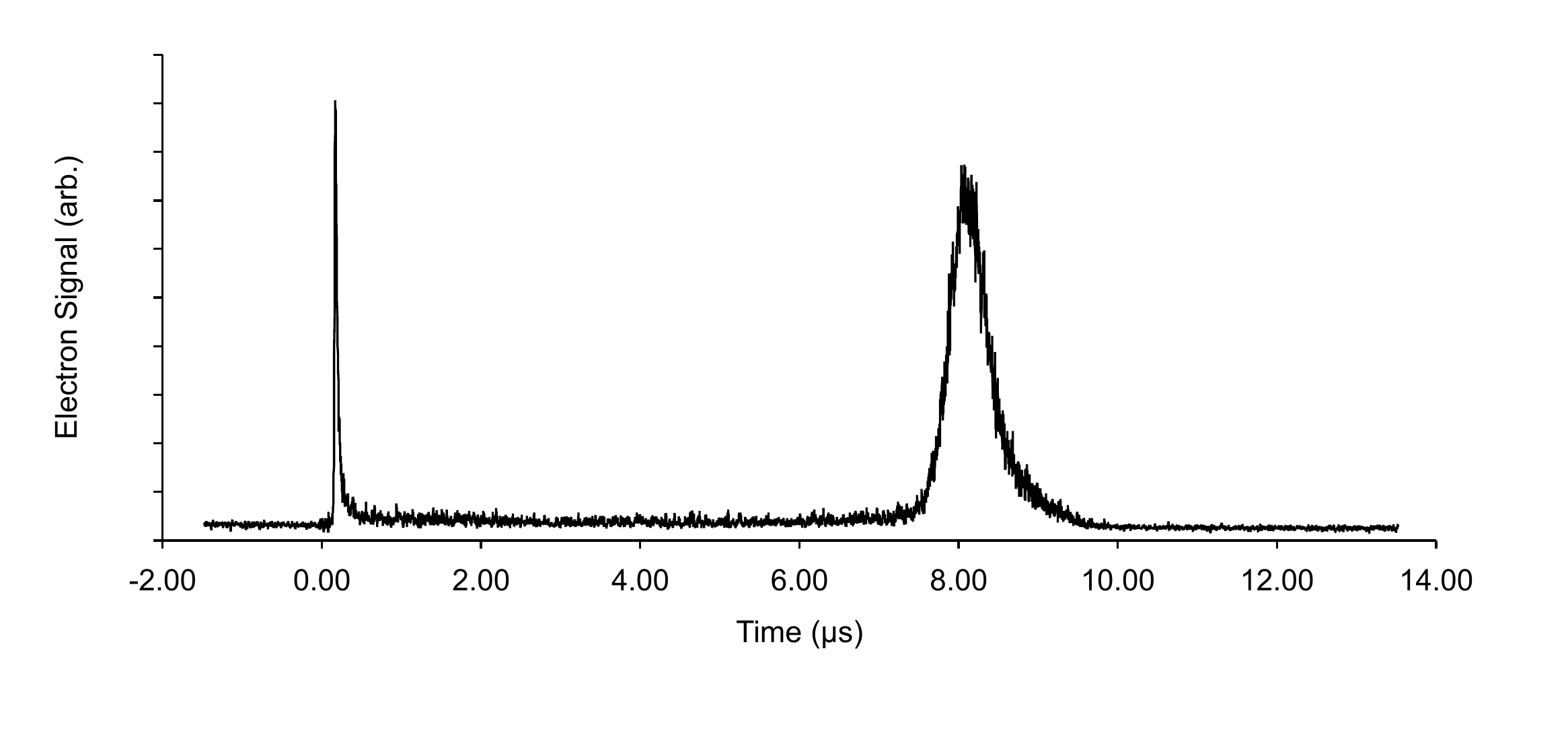}
\caption{Oscilloscope trace showing the arrival time of electrons produced following two-color production of NO molecules in the $52f(2)$ Rydberg state with -250 mV applied to G$_{1}$ and 60 V applied to G$_{3}$.}
\label{Figure_2.pdf}
\end{figure}

Excited molecules formed in this illuminated volume move with the jet through a nominally field-free region to reach an exit grid, $G_{2}$, as diagrammed in Figure~\ref{Figure_1.pdf}.  At certain frequencies of $\omega_{2}$ well below the two-color ionization threshold of NO, the combination of laser pulses at $\omega_{1}$ and $\omega_{2}$ produces an electron signal waveform exemplified by Figure~\ref{Figure_2.pdf}.  This waveform consists of a prompt electron signal, which appears at the MCP within about 200 nanoseconds of $\omega_{2}$ excitation, followed by a broader pulse of electrons that arrives 8.2 $\mu$s later.  

The shape and arrival time of the prompt pulse depends slightly on the magnitude of the very low DC field applied between grids 1 and 2.  At zero field, the prompt peak reaches the detector 250 ns after $\omega_2$.  When we bias G$_{1}$ to form a weak 250 mV acceleration field, the prompt electron signal sharpens to arrive at 170 ns, as in Figure~\ref{Figure_2.pdf}, while the delayed signal remains unaffected.  

The amplitude of the delayed electron pulse depends on the magnitude of the DC voltage applied to the flight-tube entrance grid, $G_{3}$, rising from zero with no gradient between $G_{2}$ and $G_{3}$, to plateau slightly higher than relative intensity shown for gradients above 100 V/cm.  For a given $G_{3}$ voltage, the intensity of this late signal does not vary to a large degree with Rydberg binding energy, and, most significantly, yields comparable numbers of electrons for $\omega_{1} + \omega_{2}$ energies below and above the field-ionization limit.  From this result, we can conclude that, by the time the illuminated volume reaches $G_{2}$, it has lost memory of the principal quantum number, $n$, to which it was initially excited.  We take this to indicate that the Rydberg molecules prepared by  $\omega_{1} + \omega_{2}$ excitation have evolved to form a plasma, and that this plasma has survived for the 8.2 $\mu$s required to reach $G_{2}$.  

The intensity profile of this signal reflects electrons extracted from this plasma volume as it passes through $G_{2}$, which produces a waveform reflecting the velocity of the molecular beam in the laboratory frame.  Thus, for a measured distance of 11 mm from the the laser interaction region to $G_{2}$ the peak arrival time in Figure~\ref{Figure_2.pdf} corresponds to average velocity of 1340 m/s.  This figure conforms well with the value of $\bar{V}^{\infty}_{\|}$ = 1358 m/s predicted for a 90:10 mixture of He and NO.  

The breadth of this waveform is determined by the width of our illumination volume, which is about 1 mm or 10 percent of the flight distance to G$_{2}$.  

The integrated electron signal collected as a function of $\omega_{2}$ forms a spectrum of the Rydberg series populated by second-photon absorption.  We can readily assign this spectrum to the $nf$ Rydberg series of NO converging to the $v^{+} = 0, N^{+} =2$ state of $^{1}\Sigma$ NO$^{+}$.  In the region of $\omega_{2}$ = 30,490 cm$^{-1}$, a few lines appear in the $np$ series converging to the $N^{+} = 0$ rotational level of the cation ground state.  

The Rydberg states of NO predissociate with rates of decay that vary by series according to orbital angular momentum, $l$, and along a series with principal quantum number, $n$.  For $n$ from 40 to 70, isolated-molecule lifetimes in the $nf$ series converging to $N^{+}$= 2 range from 10 to 30 ns.  Intrinsic lifetimes for $np$ states fall below 1 ns for $n$~\textless~70.  The inhomogeneous electrostatic fields presented by neighboring ions cause $l$- and $m_{l}$-mixing, which can lengthen these lifetimes appreciably \cite{Vrakking_l}.

Previous investigators have characterized these properties of the high-Rydberg states of NO by means of double-resonant laser excitation followed by pulsed-field ionization \cite{Vrakking_NO}.  In our experiment, resonant absorption is signaled by the prompt production of free electrons, in the absence of a pulsed field, at total energies as much as 80 cm$^{-1}$ (10 meV) below the lowest ionization threshold of NO.  This behavior closely parallels the emission of electrons from atomic Rydberg gases comparably prepared in optical traps \cite{Killian, Feldbaum, AWalzFlannigan, Li, Roberts, Fletcher}.  

When we apply an electrostatic pulse in our apparatus (e.g. 30 V/cm for n = 50, delayed 500 ns), we produce a field-ionization signal, but the delayed signal we associate with plasma electrons remains undiminished.  

We can reverse the polarity of the field ionization pulse with similarly no effect on this plasma signal.  In fact, we can apply delayed, reverse-bias pulses as high as 150 V/cm, and observe surprisingly little effect on the signal produced by electrons released when the illuminated volume passes $G_{2}$, even though the amplitude of such a pulse substantially exceeds the field ionization threshold for all the principal quantum numbers in this range.  

\begin{figure}
\includegraphics[width=\columnwidth]{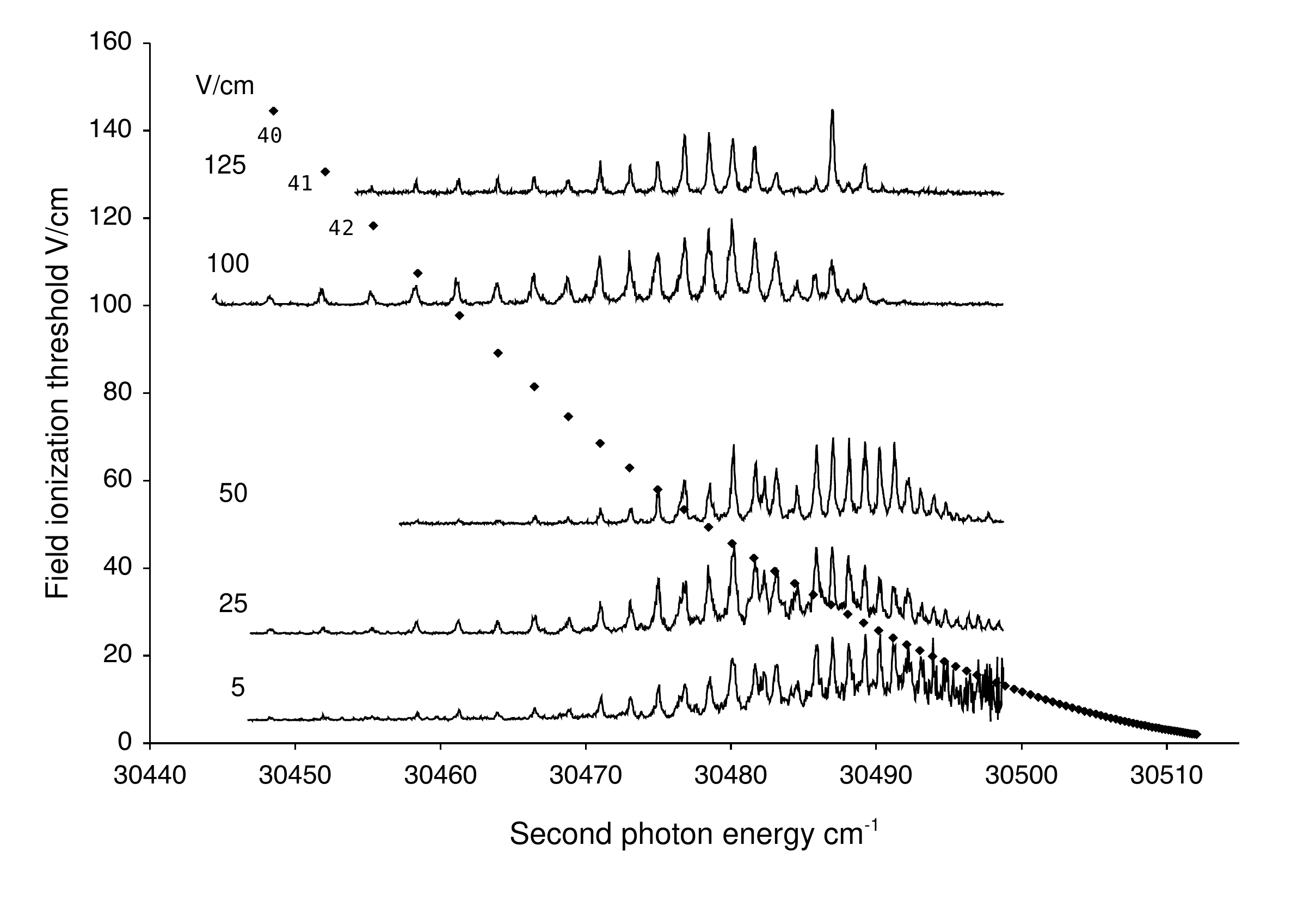}
\caption{High-Rydberg excitation spectra of the plasma signal under conditions in which 1 $\mu$s positive pulses at voltages producing the electrostatic fields indicated are applied starting 500 ns after $\omega_{2}$.  Diamonds show the adiabatic field ionization threshold for the principal quantum numbers from 40 to 110}
\label{Figure_3.pdf}
\end{figure}

Figure~\ref{Figure_3.pdf} shows a sequence of spectra obtained by collecting the plasma signal present following the application of reversed polarity field-ionization pulses with amplitudes ranging from 5 to 125 V/cm.  This experimental result again supports the notion that our Rydberg gas has evolved to form a cold plasma.  Note in particular that the electron signal extracted by the passage of the active volume into a field of 60 V cm$^{-1}$ appears undiminished even after the application of a pulsed field at more than twice this gradient.  The plasma resists destruction by an electrostatic potential that is well in excess of the field ionization threshold of the underlying Rydberg states, and then regeneratively sheds electrons in response to the application of a subsequent smaller gradient.  

We can understand this property of the plasma in terms of its predicted Debye screening length.  Assuming that Rydberg-Rydberg, Penning interactions promote electrons in one-half of the NO Rydberg states in our sample, and fewer than one percent of these electrons escape the active volume, an electron density, $n_{e}$, of 5 x 10$^{12}$ cm$^{-3}$ would remain.  Allowing for electron temperatures as high as T$_{e}$ = 100 K, we obtain a Debye screening length, $\lambda_{D}$=$\sqrt{\epsilon_{0}kT_{e}/q_{e}^{2}n_{e}}$ \textless~0.4 $\mu$m, far smaller than the width of our active volume, a point of scale marking the onset of collective effects.  Thus, the second field gradient at G$_{2}$ extracts electrons that have redistributed from the core of the plasma to its surface in the time following the reverse-bias pulse at 1 $\mu$s.  

While the plasma signal is remarkably robust to the application of a several-hundred V/cm field-ionization pulse a half-microsecond or more after $\omega_{2}$, an electrostatic pulse of much lower amplitude, applied in coincidence with $\omega_{2}$ (at t = 0) completely extinguishes the late signal.  When this pulse is applied with increasing delay after t = 0, the late signal grows, reaching its field-free amplitude after 200 ns.  

We find that both the amplitude and this pulsed-field-measured appearance time for a durable plasma signal depends on the properties of the initially prepared population of high-Rydberg NO molecules.  If the density of excited states, which can be regulated by the delay between $\omega_{1}$ and  $\omega_{2}$, is too low, we see no plasma signal.  If the temperature of NO in the molecular beam, as reflected by its rotational line intensities in the $\omega_{1}$ spectrum, is too high, we see no plasma signal.  

Energy conservation demands energy pooling for the production of free electrons and a plasma from an ensemble of NO molecules that have initial energies below the ionization threshold.  Density clearly regulates the encounter frequency by which this can occur.  

The effect of temperature is less clear.  Increased rotational temperature reduces density, because a smaller fraction of the molecules in the beam populates the rotational ground state from which double-resonant excitation originates.  

It is also possible that attractive intermolecular interactions drive disproportionating Rydberg-Rydberg electronic energy transfer in NO, analogous to the dipole-dipole coupling that ignites ultracold plasmas in cold gases of Rydberg atoms \cite{Li_dipole}.  In such circumstances, thermal translational energy could be expected to inhibit interaction.     

Finally, we return to the question of dissociative recombination.  The energy per excited NO molecule in our sample greatly exceeds the thermochemical threshold to form N $^{4}$S and O $^{3}$P.  Dissociation to neutral atoms represents a dissipation path at all stages of this experiment.  Results presented above suggest that the Rydberg gas in our system evolves to a plasma on a timescale of 200 ns.  Intrinsic predissociation lifetimes are found to be much shorter than this.  Lifetime lengthening by $l$ and $m_{l}$ mixing in the inhomogeneous fields of nearby ions must operate under our conditions to close this relaxation channel.  

A similar dynamics must apply for dissociative recombination, the rate of which is well established for NO under binary collision conditions.  This rate increases with decreasing cation-electron collision energy.  For an electron temperature of 100 K, theory and experiment predict the lifetime of a single NO$^{+}$ cation to be 700 ns at the electron density of our plasma \cite{Carata}.  At such a rate, bimolecular cation-electron recombination would extinguish the plasma long before it reached G$_{2}$.  The plasma environment, with its inhomogeneous field of incompletely shielded charges, must operate to admix high-$l$ partial waves in cation-electron scattering states, resulting in angular momentum barriers that block core penetration to inhibit dissociative recombination, much like predissociation is inhibited for discrete states.  The plasma environment has been seen directly to mix $l$ and $m_{l}$ in atomic systems, with consequences for delayed electron photoemission \cite{Raithel_ponder,Feldbaum}.  A similar many-body electrodynamics likely accounts for the durability of our ultracold molecular plasma.  

This work was supported by the Natural Sciences and Engineering Research Council of Canada (NSERC). 

\bibliography{PRL_papers}

\begin{thebibliography}{21}
\expandafter\ifx\csname natexlab\endcsname\relax\def\natexlab#1{#1}\fi
\expandafter\ifx\csname bibnamefont\endcsname\relax
  \def\bibnamefont#1{#1}\fi
\expandafter\ifx\csname bibfnamefont\endcsname\relax
  \def\bibfnamefont#1{#1}\fi
\expandafter\ifx\csname citenamefont\endcsname\relax
  \def\citenamefont#1{#1}\fi
\expandafter\ifx\csname url\endcsname\relax
  \def\url#1{\texttt{#1}}\fi
\expandafter\ifx\csname urlprefix\endcsname\relax\def\urlprefix{URL }\fi
\providecommand{\bibinfo}[2]{#2}
\providecommand{\eprint}[2][]{\url{#2}}

\bibitem[{\citenamefont{Ichimaru}(1982)}]{Ichimaru}
\bibinfo{author}{\bibfnamefont{S.}~\bibnamefont{Ichimaru}},
  \bibinfo{journal}{Reviews of Modern Physics} \textbf{\bibinfo{volume}{54}},
  \bibinfo{pages}{1017} (\bibinfo{year}{1982}).

\bibitem[{\citenamefont{Killian et~al.}(2007)\citenamefont{Killian, Pattard,
  Pohl, and Rost}}]{Killian:2008}
\bibinfo{author}{\bibfnamefont{T.}~\bibnamefont{Killian}},
  \bibinfo{author}{\bibfnamefont{T.}~\bibnamefont{Pattard}},
  \bibinfo{author}{\bibfnamefont{T.}~\bibnamefont{Pohl}}, \bibnamefont{and}
  \bibinfo{author}{\bibfnamefont{J.}~\bibnamefont{Rost}},
  \bibinfo{journal}{Physics Reports} \textbf{\bibinfo{volume}{449}},
  \bibinfo{pages}{77} (\bibinfo{year}{2007}).

\bibitem[{\citenamefont{Amthor et~al.}(2007)\citenamefont{Amthor, Reetz-Lamour,
  Giese, and Weidem{\"u}ller}}]{Weidemuller_many}
\bibinfo{author}{\bibfnamefont{T.}~\bibnamefont{Amthor}},
  \bibinfo{author}{\bibfnamefont{M.}~\bibnamefont{Reetz-Lamour}},
  \bibinfo{author}{\bibfnamefont{C.}~\bibnamefont{Giese}}, \bibnamefont{and}
  \bibinfo{author}{\bibfnamefont{M.}~\bibnamefont{Weidem{\"u}ller}},
  \bibinfo{journal}{Phys. Rev. A} \textbf{\bibinfo{volume}{76}},
  \bibinfo{pages}{054702} (\bibinfo{year}{2007}).

\bibitem[{\citenamefont{Robicheaux et~al.}(2004)\citenamefont{Robicheaux,
  Hern{\'a}ndez, Top{\c c}u, and Noordam}}]{Robicheaux_coherent}
\bibinfo{author}{\bibfnamefont{F.}~\bibnamefont{Robicheaux}},
  \bibinfo{author}{\bibfnamefont{J.~V.} \bibnamefont{Hern{\'a}ndez}},
  \bibinfo{author}{\bibfnamefont{T.}~\bibnamefont{Top{\c c}u}},
  \bibnamefont{and} \bibinfo{author}{\bibfnamefont{L.~D.}
  \bibnamefont{Noordam}}, \bibinfo{journal}{Phys. Rev. A}
  \textbf{\bibinfo{volume}{70}}, \bibinfo{pages}{042703}
  (\bibinfo{year}{2004}).

\bibitem[{\citenamefont{Murillo}(2007)}]{Murillo}
\bibinfo{author}{\bibfnamefont{M.~S.} \bibnamefont{Murillo}},
  \bibinfo{journal}{Phys. Plasmas} \textbf{\bibinfo{volume}{14}},
  \bibinfo{pages}{055702} (\bibinfo{year}{2007}).

\bibitem[{\citenamefont{Ates et~al.}(2007)\citenamefont{Ates, Pohl, Pattard,
  and Rost}}]{Ates}
\bibinfo{author}{\bibfnamefont{C.}~\bibnamefont{Ates}},
  \bibinfo{author}{\bibfnamefont{T.}~\bibnamefont{Pohl}},
  \bibinfo{author}{\bibfnamefont{T.}~\bibnamefont{Pattard}}, \bibnamefont{and}
  \bibinfo{author}{\bibfnamefont{J.~M.} \bibnamefont{Rost}},
  \bibinfo{journal}{Phys. Rev. A} \textbf{\bibinfo{volume}{76}},
  \bibinfo{pages}{013413} (\bibinfo{year}{2007}).

\bibitem[{\citenamefont{Knuffman and Raithel}(2007)}]{Raithel_ponder}
\bibinfo{author}{\bibfnamefont{B.}~\bibnamefont{Knuffman}} \bibnamefont{and}
  \bibinfo{author}{\bibfnamefont{G.}~\bibnamefont{Raithel}},
  \bibinfo{journal}{Phys. Rev. A} \textbf{\bibinfo{volume}{75}},
  \bibinfo{pages}{053401} (\bibinfo{year}{2007}).

\bibitem[{\citenamefont{Tanner et~al.}(2008)\citenamefont{Tanner, Han, Shuman,
  and Gallagher}}]{Tanner}
\bibinfo{author}{\bibfnamefont{P.~J.} \bibnamefont{Tanner}},
  \bibinfo{author}{\bibfnamefont{J.}~\bibnamefont{Han}},
  \bibinfo{author}{\bibfnamefont{E.~S.} \bibnamefont{Shuman}},
  \bibnamefont{and} \bibinfo{author}{\bibfnamefont{T.~F.}
  \bibnamefont{Gallagher}}, \bibinfo{journal}{Phys. Rev. Lett.}
  \textbf{\bibinfo{volume}{100}}, \bibinfo{pages}{043002}
  (\bibinfo{year}{2008}).

\bibitem[{\citenamefont{Killian et~al.}(1999)\citenamefont{Killian, Kulin,
  Bergeson, Orozco, Orzel, and Rolston}}]{Killian}
\bibinfo{author}{\bibfnamefont{T.~C.} \bibnamefont{Killian}},
  \bibinfo{author}{\bibfnamefont{S.}~\bibnamefont{Kulin}},
  \bibinfo{author}{\bibfnamefont{S.~D.} \bibnamefont{Bergeson}},
  \bibinfo{author}{\bibfnamefont{L.~A.} \bibnamefont{Orozco}},
  \bibinfo{author}{\bibfnamefont{C.}~\bibnamefont{Orzel}}, \bibnamefont{and}
  \bibinfo{author}{\bibfnamefont{S.~L.} \bibnamefont{Rolston}},
  \bibinfo{journal}{Phys. Rev. Lett.} \textbf{\bibinfo{volume}{83}},
  \bibinfo{pages}{4776} (\bibinfo{year}{1999}).

\bibitem[{\citenamefont{Feldbaum et~al.}(2002)\citenamefont{Feldbaum, Morrow,
  Dutta, and Raithel}}]{Feldbaum}
\bibinfo{author}{\bibfnamefont{D.}~\bibnamefont{Feldbaum}},
  \bibinfo{author}{\bibfnamefont{N.~V.} \bibnamefont{Morrow}},
  \bibinfo{author}{\bibfnamefont{S.~K.} \bibnamefont{Dutta}}, \bibnamefont{and}
  \bibinfo{author}{\bibfnamefont{G.}~\bibnamefont{Raithel}},
  \bibinfo{journal}{Phys. Rev. Lett.} \textbf{\bibinfo{volume}{89}},
  \bibinfo{pages}{173004} (\bibinfo{year}{2002}).

\bibitem[{\citenamefont{Walz-Flannigan
  et~al.}(2004)\citenamefont{Walz-Flannigan, Guest, Choi, and
  Raithel}}]{AWalzFlannigan}
\bibinfo{author}{\bibfnamefont{A.}~\bibnamefont{Walz-Flannigan}},
  \bibinfo{author}{\bibfnamefont{J.~R.} \bibnamefont{Guest}},
  \bibinfo{author}{\bibfnamefont{J.-H.} \bibnamefont{Choi}}, \bibnamefont{and}
  \bibinfo{author}{\bibfnamefont{G.}~\bibnamefont{Raithel}},
  \bibinfo{journal}{Phys. Rev. A} \textbf{\bibinfo{volume}{69}},
  \bibinfo{pages}{063405} (\bibinfo{year}{2004}).

\bibitem[{\citenamefont{Li et~al.}(2004)\citenamefont{Li, Noel, Robinson,
  Tanner, Gallagher, , Comparat, LaburtheTolra, Vanhaecke, Vogt et~al.}}]{Li}
\bibinfo{author}{\bibfnamefont{W.}~\bibnamefont{Li}},
  \bibinfo{author}{\bibfnamefont{M.~W.} \bibnamefont{Noel}},
  \bibinfo{author}{\bibfnamefont{M.~P.} \bibnamefont{Robinson}},
  \bibinfo{author}{\bibfnamefont{P.~J.} \bibnamefont{Tanner}},
  \bibinfo{author}{\bibfnamefont{T.~F.} \bibnamefont{Gallagher}}, ,
  \bibinfo{author}{\bibfnamefont{D.}~\bibnamefont{Comparat}},
  \bibinfo{author}{\bibfnamefont{B.}~\bibnamefont{LaburtheTolra}},
  \bibinfo{author}{\bibfnamefont{N.}~\bibnamefont{Vanhaecke}},
  \bibinfo{author}{\bibfnamefont{T.}~\bibnamefont{Vogt}}, \bibnamefont{et~al.},
  \bibinfo{journal}{Phys. Rev. A} \textbf{\bibinfo{volume}{70}},
  \bibinfo{pages}{042713} (\bibinfo{year}{2004}).

\bibitem[{\citenamefont{Roberts et~al.}(2004)\citenamefont{Roberts, Fertig,
  Lim, and Rolston}}]{Roberts}
\bibinfo{author}{\bibfnamefont{J.~L.} \bibnamefont{Roberts}},
  \bibinfo{author}{\bibfnamefont{C.~D.} \bibnamefont{Fertig}},
  \bibinfo{author}{\bibfnamefont{M.~J.} \bibnamefont{Lim}}, \bibnamefont{and}
  \bibinfo{author}{\bibfnamefont{S.~L.} \bibnamefont{Rolston}},
  \bibinfo{journal}{Phys. Rev. Lett.} \textbf{\bibinfo{volume}{92}},
  \bibinfo{pages}{253003} (\bibinfo{year}{2004}).

\bibitem[{\citenamefont{Fletcher et~al.}(2007)\citenamefont{Fletcher, Zhang,
  and Rolston}}]{Fletcher}
\bibinfo{author}{\bibfnamefont{R.~S.} \bibnamefont{Fletcher}},
  \bibinfo{author}{\bibfnamefont{X.~L.} \bibnamefont{Zhang}}, \bibnamefont{and}
  \bibinfo{author}{\bibfnamefont{S.~L.} \bibnamefont{Rolston}},
  \bibinfo{journal}{Phys. Rev. Lett.} \textbf{\bibinfo{volume}{99}},
  \bibinfo{pages}{145001} (\bibinfo{year}{2007}).

\bibitem[{\citenamefont{Vitrant et~al.}(1982)\citenamefont{Vitrant, Raimond,
  Gross, and Haroche}}]{Vitrant}
\bibinfo{author}{\bibfnamefont{G.}~\bibnamefont{Vitrant}},
  \bibinfo{author}{\bibfnamefont{J.}~\bibnamefont{Raimond}},
  \bibinfo{author}{\bibfnamefont{M.}~\bibnamefont{Gross}}, \bibnamefont{and}
  \bibinfo{author}{\bibfnamefont{S.}~\bibnamefont{Haroche}},
  \bibinfo{journal}{J. Phys. B} \textbf{\bibinfo{volume}{15}},
  \bibinfo{pages}{L49} (\bibinfo{year}{1982}).

\bibitem[{\citenamefont{Miller}(1988)}]{Miller}
\bibinfo{author}{\bibfnamefont{D.~R.} \bibnamefont{Miller}}, in
  \emph{\bibinfo{booktitle}{Atomic and Molecular Beam Methods}}, edited by
  \bibinfo{editor}{\bibfnamefont{G.}~\bibnamefont{Scoles}}
  (\bibinfo{publisher}{Oxford University Press}, \bibinfo{address}{New York},
  \bibinfo{year}{1988}), pp. \bibinfo{pages}{14--53}.

\bibitem[{\citenamefont{Beijerinck and Verster}(1981)}]{Beijerinck}
\bibinfo{author}{\bibfnamefont{H.~C.~W.} \bibnamefont{Beijerinck}}
  \bibnamefont{and} \bibinfo{author}{\bibfnamefont{N.~F.}
  \bibnamefont{Verster}}, \bibinfo{journal}{Physica B+C}
  \textbf{\bibinfo{volume}{111C}}, \bibinfo{pages}{327} (\bibinfo{year}{1981}).

\bibitem[{\citenamefont{Vrakking and Lee}(1995{\natexlab{a}})}]{Vrakking_l}
\bibinfo{author}{\bibfnamefont{M.}~\bibnamefont{Vrakking}} \bibnamefont{and}
  \bibinfo{author}{\bibfnamefont{Y.}~\bibnamefont{Lee}},
  \bibinfo{journal}{Phys. Rev. A} \textbf{\bibinfo{volume}{51}},
  \bibinfo{pages}{R894} (\bibinfo{year}{1995}{\natexlab{a}}).

\bibitem[{\citenamefont{Vrakking and Lee}(1995{\natexlab{b}})}]{Vrakking_NO}
\bibinfo{author}{\bibfnamefont{M.}~\bibnamefont{Vrakking}} \bibnamefont{and}
  \bibinfo{author}{\bibfnamefont{Y.}~\bibnamefont{Lee}}, \bibinfo{journal}{J.
  Chem. Phys} \textbf{\bibinfo{volume}{102}}, \bibinfo{pages}{8818}
  (\bibinfo{year}{1995}{\natexlab{b}}).

\bibitem[{\citenamefont{Li et~al.}(2005)\citenamefont{Li, Tanner, and
  Gallagher}}]{Li_dipole}
\bibinfo{author}{\bibfnamefont{W.}~\bibnamefont{Li}},
  \bibinfo{author}{\bibfnamefont{P.~J.} \bibnamefont{Tanner}},
  \bibnamefont{and} \bibinfo{author}{\bibfnamefont{T.~F.}
  \bibnamefont{Gallagher}}, \bibinfo{journal}{Phys. Rev. Lett.}
  \textbf{\bibinfo{volume}{94}}, \bibinfo{pages}{173001}
  (\bibinfo{year}{2005}).

\bibitem[{\citenamefont{Schneider et~al.}(2000)\citenamefont{Schneider,
  Rabad{\'a}n, Carata, Andersen, Suzor-Weiner, and Tennyson}}]{Carata}
\bibinfo{author}{\bibfnamefont{I.~F.} \bibnamefont{Schneider}},
  \bibinfo{author}{\bibfnamefont{I.}~\bibnamefont{Rabad{\'a}n}},
  \bibinfo{author}{\bibfnamefont{L.}~\bibnamefont{Carata}},
  \bibinfo{author}{\bibfnamefont{L.~H.} \bibnamefont{Andersen}},
  \bibinfo{author}{\bibfnamefont{A.}~\bibnamefont{Suzor-Weiner}},
  \bibnamefont{and} \bibinfo{author}{\bibfnamefont{J.}~\bibnamefont{Tennyson}},
  \bibinfo{journal}{J. Phys. B} \textbf{\bibinfo{volume}{33}},
  \bibinfo{pages}{4849} (\bibinfo{year}{2000}).

\end{thebibliography}

\end{document}